\documentclass[prl,twocolumn,superscriptaddress]{revtex4}

\usepackage{amsmath,amsfonts,amssymb,bbm}
\usepackage{graphicx}

\usepackage{color}
\usepackage[pdfpagelabels,pdftex,bookmarks,breaklinks]{hyperref}
\definecolor{dullmagenta}{RGB}{100,0,120} % choose colors
\definecolor{darkviolet}{RGB}{140,0,160}
\definecolor{darkgreen}{RGB}{0,150,0}
\definecolor{darkblue}{RGB}{0,0,150}
\hypersetup{colorlinks, linkcolor=dullmagenta, citecolor=darkgreen, filecolor=darkblue, urlcolor=darkviolet}
\hypersetup{pdftitle=Heralded\ Polynomial-Time\ Quantum\ State\ Tomography}% add a title to the metadata

\def\>{\rangle}
\def\<{\langle}
\def\({\left(}
\def\){\right)}

\newcommand{\ket}[1]{|#1\>}
\newcommand{\bra}[1]{\<#1|}

\DeclareMathOperator{\Tr}{Tr}
\DeclareMathOperator{\tr}{tr}

\begin{document}

\title{Heralded Polynomial-Time Quantum State Tomography}

\author{Steven T. Flammia}
\affiliation{Perimeter Institute for Theoretical Physics, Waterloo, Ontario, N2L 2Y5 Canada}

\author{David Gross}
\affiliation{Institute for Theoretical Physics, Leibniz University Hannover, 30167 Hannover, Germany}

\author{Stephen D. Bartlett}
\affiliation{School of Physics, The University of Sydney, Sydney, New South Wales 2006, Australia}

\author{Rolando Somma}
\affiliation{Perimeter Institute for Theoretical Physics, Waterloo, Ontario, N2L 2Y5 Canada}

\date{20 February 2010}

\begin{abstract} 
We describe an algorithm for quantum state tomography that converges
in polynomial time to an estimate, together with a rigorous error
bound on the fidelity between the estimate and the true state. The
result suggests that state tomography on large quantum systems may be
much more feasible than the exponential size of state space
suggests. In many situations, the correctness of the state estimate
can be certified from the data alone, with no {\it a priori\/} assumptions on the
form of the measured state.
%The statement continues to
%hold in some situations, even in the abscence of a priori assumptions
%on the form of the encountered states. 
%For states that are well approximated by a matrix product state
%with small bond dimension, the algorithm outputs an estimate which is
%of size polynomial in the bond dimension and linear in the number of
%qubits. When no such matrix product state exists, the algorithm
%heralds an error.  
\end{abstract}

\maketitle
% -----------------------------------------------------------------------------------------------------------%

Quantum state tomography (QST) --- the process of estimating a quantum
state using local measurements given a large number of copies --- is
an exceptionally daunting task.  As it is typically formulated
\cite{Paris2004}, simply to output an estimate for a generic state of
$n$ qubits would take exponential time in $n$, given that there are an
exponential number of coefficients in a generic state's description.
This is but one of several inefficiencies. Most quantum states have
exponentially small amplitudes in almost every basis, so to
distinguish any one of these amplitudes from zero takes exponential
time.  Assuming one were able to collect all of that data from an
informationally complete measurement,
% which necessarily has an
%exponential number of outcomes, 
one is left with the intractable computational task of inverting the
measured frequencies to find an estimate of the state.

These barriers to performing efficient quantum tomography for generic
states are not necessarily applicable to real-world situations,
however, as the traditional representation of quantum states in terms
of Hilbert space vectors and density matrices is in a sense too
general. Indeed, states which occur in many practical situations are
specified by a small number of parameters. An efficient description
could be a polynomial-sized quantum circuit which outputs the state;
or, in the case of thermodynamical equilibrium states, a
bounded-interaction-range Hamiltonian and a temperature.  This insight
is not new: researchers in many-body physics and quantum information
theory have found many classes of states which are described by a
number of parameters scaling polynomially in $n$ and which closely
approximate the kind of states found in particular physical
systems~\cite{Fannes1992, Hastings2006, Hastings2007a,
Perez-Garcia2007a, Schuch2008a, Verstraete2006b, Vidal2003a}.

However, the question of whether these restricted classes can be put
to use in a tomographic setting has remained largely open. Any result
closing this gap should address three points: \emph{1.} Choose a
few-parameter set of states adapted to the physical nature of the
system under consideration.  \emph{2.} Specify a protocol which
efficiently \footnote{Throughout, ``efficient'' means that both the
number of measurements and the classical post-processing time scale
polynomially in the number of qubits.} identifies these parameters
from a small and simple-to-implement set of measurements. \emph{3.} If
possible, give an independent way of verifying that the protocol
produced a faithful estimate of the physical state (without having to
\emph{assume} that the set chosen in the first step correctly
describes the system).

We briefly explain the third point. An exprimentalist should be wary of any tomographic procedure for which the success is predicated on some abstract technical condition (e.g., the system is well-described by a ``matrix product state of bond dimension $D$'').  Therefore, it
is desirable that the protocol infers from the data alone whether or
not it can guarantee a faithful reconstruction, and aborts if it
cannot. We say that such a procedure \emph{heralds} its own success.

%Furthermore, efficient certification
%that a given state satisfies an {\it a priori\/} promise (e.g.\ ``it
%is the output of some polynomial-sized quantum circuit'' or ``it is a
%Gibbs state of some Hamiltonian at some temperature'') should also be
%investigated.  

In this work, we address all three problems above by presenting an
algorithm for heralded quantum state tomography using a polynomial
amount (poly($n$)) of measurement and classical computation.  

The physical system we have in mind is one where the constituents are
arranged in a one-dimensional configuration (e.g. ions in a linear
trap \cite{Haffner2005}).
It is highly plausible that in such a setting,
correlations between neighboring qubits are much more pronounced (due
to direct interaction) than correlations between distant systems
(mediated e.g.\ by global fluctuations of control fields). A
polynomially-sized class of states anticipating exactly this behavior
has long been studied under the names of finitely correlated states
(FCS) or matrix product states (MPS)
\cite{Fannes1992,Perez-Garcia2007}. Below, we present the
\emph{heralded MPS tomography} (H-MPS) algorithm. Given experimental
data, it outputs either a polynomial-sized pure matrix product state
together with a rigorous bound on the fidelity between the true
density matrix and the estimate, or else it returns a failure flag.
This happens in particular when such an efficient description does not
exist. At least in an idealized scenario, one can \emph{guarantee}
that an efficient MPS description will be found in polynomial time,
given that it exists.

In later sections, we sketch generalizations to higher dimensions, 
to highly mixed systems, and to quantum channels describing correlated
noise. We emphasize that the H-MPS algorithm can be employed
whenever one expects that an MPS description for a system may be
appropriate, whether or not it possesses a one-dimensional geometry.
Examples of MPS include the GHZ, W, cluster, and AKLT states. Our
results also address an open question of Aaronson~\cite{Aaronson2007}
by demonstrating that MPS can be learned efficiently.

To find an unknown quantum state $\rho$ of $n$ qudits ($d$-dimensional
quantum systems), we propose the following procedure. One selects a
number $k\ll n$, such that obtaining an estimate for the state of $2k$
contiguous qudits using standard tomographic techniques
\cite{Paris2004} is
feasible. We denote the reduction of $\rho$ to sites $j, j+1, \ldots,
j+(2k-1)$ by $\rho_j$, and we use $\sigma_{j}$ to refer to a
tomographic estimate of this state. Assume that by standard
methods one obtains a confidence interval such that, for all $j$, the trace norm satisfies
\begin{equation}\label{eqn:confidence}
	\|\rho_j - \sigma_j\|_{\tr} \leq \epsilon_j
\end{equation}
with high confidence.

The data $\{\sigma_j, \epsilon_j\}$  are then passed to the H-MPS
algorithm. It either declares a failure and aborts, or else outputs an efficiently
describable state $\ket\Psi$ and a number $\tau$ such that
the fidelity provably obeys
\begin{equation}\label{eqn:fidelity}
	F(\ket\Psi, \rho) = \sqrt{\bra\Psi\rho\ket\Psi} \geq \sqrt{1 -
	\tau}
\end{equation}
whenever (\ref{eqn:confidence}) holds. 

It may come as a surprise that such a certificate is possible. After
all, there are exponentially many degrees of freedom which remain
unmeasured in the procedure. How can we guarantee that there are no
two orthogonal states which happen to coincide on the tiny number of
coefficients we have obtained information about? Geometrically, the
reason is that a generic MPS is ``locally exposed'' \cite{Fannes1992},
i.e.\ it lies at an extreme point of the space of quantum states,
whose tangent plane corresponds to a local functional. In physical
terms: with every MPS, one can associate a gapped, local Hamiltonian,
which acts as a witness for that state \cite{Fannes1992,Perez-Garcia2007}. 
The main technical issue solved below is finding a simple way of bounding 
this gap for finite, non-translationally invariant and possibly degenerate systems.

Before describing the algorithm, we pause to recall the definition of
MPS \cite{Fannes1992,Perez-Garcia2007}.  A state vector $\ket\Psi$ on
$n$ qudits is an \emph{MPS with bond-dimension $D$} if there are 
$n d$ complex $D\times D$ matrices
$\{ A_j^s\,\},\, s=1,\ldots,d; j = 1,\ldots,n$ such that
\begin{equation}\label{eqn:mps}
  |\Psi\rangle = \sum_{s_1,\ldots,s_n=1}^d
  {\rm Tr}[A_1^{s_1}A_2^{s_2}\cdots
  A_n^{s_n}]|s_1,s_2,\ldots,s_n\rangle \,.
\end{equation}
Here, $D$ is a free parameter which describes the complexity of the
model. It is well-known that any state has an MPS description if one
allows $D$ to scale exponentially with $n$.  However, the motivation
for the seemingly ad-hoc definition (\ref{eqn:mps}) stems from the
fact that many natural states are well-approximate by an MPS with small $D$.
Note there are only $nD^2$ parameters specifying $\ket\Psi$.

%(Actually, fewer if one removes certain gauge freedom.)  
%Any quantum state can be
%represented as a MPS, but in general $D$ will depend exponentially on
%$n$.  Fortunately, MPS provide efficient descriptions of many
%interesting quantum states when $D$ is constant or grows only
%polynomially in $n$.

%Consider a one-dimensional graph
%where the edges correspond to maximally entangled states
%$\sum_{\mu=1}^D \ket{\mu,\mu}$ of pairs of $D$-dimensional systems,
%one at each site associated with this edge.
%%; see Fig.~\ref{F:MPS}.  
%Here, $D$ is a free parameter in our description.  A MPS is defined by
%mapping pairs of systems at each site $j$ to a subspace of dimension
%$d$ via a partial isometry $A_j:\C^{D^2} \rightarrow \C^d$.
%Expressing this in a basis as $A_j = \sum_{\mu,\nu=1}^D \sum_{s=1}^d
%%(A_j^s)_{\mu\nu}|s\rangle\langle\mu,\nu|$, 
%the set of matrices $\{
%A_j^s\,, s=1,\ldots,d\}$ for each site $j$ define the MPS
%$|\Psi\rangle$ by
%\begin{equation}
%  |\Psi\rangle = \sum_{s_1,\ldots,s_n=1}^d
%  {\rm Tr}[A_1^{s_1}A_2^{s_2}\cdots
%  A_n^{s_n}]|s_1,s_2,\ldots,s_n\rangle \,.
%\end{equation}
%Note there are $nD^2$ parameters specifying such a MPS.  
%%(Actually, fewer if one removes certain gauge freedom.)  
%Any quantum state can be
%represented as a MPS, but in general $D$ will depend exponentially on
%$n$.  Fortunately, MPS provide efficient descriptions of many
%interesting quantum states when $D$ is constant or grows only
%polynomially in $n$.

\paragraph{H-MPS.} The algorithm consists of two steps. In the first
stage, we compute an MPS estimate from the experimental data. The task
of the second step is to certify the fidelity of the estimate.
Because we will retrospectively verify the output of step one, it is
not necessary to insist on theoretical guarantees on the performance
of this part. Any ansatz, even an ``educated guess'', can lead to a
rigorous result, as long as it will be vindicated by passing the
second stage of H-MPS. Therefore, we give a list of algorithms for
finding MPS estimates below, without definitely endorsing one over the
others. A more detailed study of practical performances will be
provided elsewhere. For clarity of presentation, we will restrict
attention to ``generic'' states at first, treating singular situations
in the end.

\emph{The Variational Method.} A natural way of obtaining FCS
estimates from data is by way of a variational algorithm: note that
the local reductions of an MPS can be computed efficiently. Therefore,
it is feasible to change the matrices one at a time, so as to minimize
the sum of the trace-norm differences between the measured estimates
and the computed local reductions. Since the target function is
bounded from below and reduced in every step, the algorithm will
certainly converge. The advantage of this method is that all the
available information is utilized, and that it is inherently stable
against small noise. On the negative side, it may be prone to run into
local minima. (We emphasize, however, that the algorithm should not be
judged by the high standards of the variational methods employed in
many-body physics, which reliably find the global minimum of energy
functions on tens of thousands of spins. A density operator even on
ten three-level ``qutrits'' depends on 3.4 billion coefficients. If
the trace-norm distance can be reliably minimized for such
comparatively tiny systems, it would make tomography possible in otherwise
intractable regimes -- even if the algorithms fails to scale to the
orders of magnitude which are standard today in numerical
condensed matter physics).

\emph{The Quantum 2-SAT Method.} Recent developments in the context of
the ``quantum satisfiability problem'' show that it is possible to
directly and efficiently find the MPS description of the ground state
space of a one-dimensional frustration-free Hamiltonian, provided that
the following condition holds: for every $1\leq j \leq n$, the
degeneracy of the restriction of the Hamiltonian to sites $1$ to $j$
does not exceed a uniform bound.  A qubit version described in
\cite{Bravyi2006} can be generalized to arbitrary situations
\cite{Osborne2009}. It easily follows that -- at least in the
idealized scenario of vanishing experimental errors -- reconstructing
an MPS description given only information about the support of the
reduced density operators (on a sufficiently large neighborhood) is
\emph{provably} efficiently possible.  While it is unclear whether
this approach is well-suited for practical purposes, it shows that
there are no computationally hard instances which would make any
attempt of finding a general solution futile from the outset.  This
may be a surprising fact, given that hard instances are known to occur
for more general frustration-free Hamiltonians with unique MPS
ground-states \cite{Schuch2008}.

\emph{The DMRG Method.} An MPS approximation with bond dimension $D$
is only possible when the reductions $\rho_j$ are mainly supported on
a space of dimension not larger than $D^2$. Denote the projection
operator on the space of the lowest $(d^{2k}-D^2)$ eigenvalues of
$\sigma_j$ by $h_j$. One can use established methods like DMRG to find
the ground state of this ``empirical parent Hamiltonian'' (see below).
Under suitable regularity conditions and assuming low noise levels,
this procedure is guaranteed to yield good estimates, as will become
clear later. Advantage: algorithms like DMRG perform extremely well
under practical conditions. Experience shows that they are fairly
immune against local minima. The disadvantage is that only some parts
of the measured information enters the procedure.

\emph{The Educated Guess.} The scenario most often encountered in current experiments~\cite{Haffner2005} is verifying the quality of a state preparation.  
Since we present a way of verifying correctness of a proposed solution, it makes sense to test a candidate MPS which did not come from measured data, but rather from the knowledge of which state one intended to create. This version of the problem
would be more aptly described as \emph{state certification\/}: simply use the MPS description of the target state to test the certificate and get a lower bound on the fidelity of the state preparation.  

We proceed to the second stage. (The presentation will necessarily
become somewhat more technical in the next paragraphs.)
%summarize the important points below.) 
The estimate obtained in the
first step is assumed to be given in MPS form (\ref{eqn:mps}). For
simplicity of language, we group blocks of $k$ contiguous qudits into
a single site. In other words, we assume without loss of generality
that $k=1$ and that the individual systems are $d^k$-dimensional. (We
will always assume that $d^k>D$, since otherwise there are more
unknown parameters in (\ref{eqn:mps}) than were measured.) To avoid
numerically ill-conditioned inputs, we bring the MPS estimate into its
standard form fulfilling $\sum_{s} A_j^s (A_j^s)^\dagger =
\mathbbm{1}$, which is always possible by a sequence of singular value
decompositions (SVD) \cite{Fannes1992,Perez-Garcia2007}. Next, compute
the SVD of the maps
\begin{equation}\label{eqn:gamma}
	\Gamma_j: X \mapsto \sum_{s_j,s_{j+1}=1}^d \Tr[X A^{s_j}_j
	A^{s_{j+1}}_{j+1}]\, \ket{s_j,s_{j+1}}.
\end{equation}
Generically, the singular values of the $\Gamma_j$'s
will be bounded away from zero \cite{Fannes1992,Perez-Garcia2007}.
Should that not be the case, we declare an error and abort (this
situation is discussed in more detail below). If the estimate passed
this first test, we are ready to construct and analyze its ``parent
Hamiltonian'' \cite{Fannes1992,Perez-Garcia2007}, which will serve as
a witness operator, certifying
closeness of the estimate to the physical state. To that end, for
every $j\in\{1,\ldots,n-1\}$, compute the reduced density matrix of
the estimate $\ket\Psi$ on sites $j, j+1$. Let $h_j$ be the
projection operator onto the kernel of this density matrix. For every
$j\in \{1,\ldots,n-1\}$ compute the eigenvalues of $h_j h_{j+1} h_j$.
Let $\gamma_j$ be the square-root of the largest eigenvalue not equal
to one. Set $\gamma=\max_j \gamma_j$.  The H-MPS algorithm concludes
by outputting $\ket\Psi$ and
$\tau=\frac1{1-2\gamma}\sum_j(\Tr(h_j\sigma_j) + \epsilon_j)$.

In the following paragraph we prove the validity of
Eq.~(\ref{eqn:fidelity}). The technical hurdle to overcome is to
lower-bound the gap in the spectrum of $H=\sum_{j=1}^{n-1} h_j$. For
non-degenerate translationally invariant (TI) systems, a (fairly
involved) proof was given in \cite{Fannes1992}.  Partly building on
their ideas, we present a short argument below, which applies to the
relevant case of finite, non-TI quantum states.

By construction, $H\geq 0, H\psi=0$, so that the lowest eigenvalue of $H$ is
zero. We claim that the second-smallest eigenvalue
$\operatorname{gap}(H)$ is lower-bounded by $(1-2\gamma)$. To prove
the assertion, we start by following \cite{Fannes1992} and note that
$\operatorname{gap}(H)=\max \{\lambda \,|\, H^2 \geq \lambda H\}$.
Compute:
\begin{eqnarray}
	H^2 &=& \sum_{j=1}^{n-1} h_j^2 + \sum_{j=1}^{n-2} \left(h_j h_{j+1} +
	h_{j+1} h_j\right) + \sum_{|j-j'|>1} h_j h_{j'} \nonumber \\
	&\geq& H + \sum_{j=1}^{n-2} [h_j, h_{j+1}]_+, \label{eqn:hsquared}
\end{eqnarray}
where we used the fact that the $h_j$ are projections and 
omitted non-negative summands. Next, we will establish
\begin{equation}\label{eqn:anticomm}
	[h_j,h_{j+1}]_+ \geq -\gamma_j (h_j + h_{j+1}).
\end{equation}
We borrow some facts
from basic Hilbert space theory \cite{Halmos1969,Avron1994}: any two
projection
operators $h_j,h_{j+1} $ can be brought simultaneously into block
diagonal form,
where on the $i$th block, $h_j$ and $h_{j+1}$ act like one-dimensional
projections whose ranges are vectors enclosing the
\emph{canonical angle} $\theta_i$. One verifies that $\cos^2\theta_i$
is the $i$th eigenvalue of $h_j h_{j+1} h_j$. Certainly
(\ref{eqn:anticomm}) holds on the entire space
if and only if that relation is true on every block.
Thus, we have reduced the problem to the case of one-dimensional
projectors in a two-dimensional space. Here, we can solve it by
elementary means and find that (\ref{eqn:anticomm}) holds for
non-trivial angles as long as
$\gamma\geq \cos\theta_i$, which is true by construction.
Plugging (\ref{eqn:anticomm}) into (\ref{eqn:hsquared}) gives
\begin{equation*}
	H^2 \geq H -\gamma\sum_j (h_j + h_{j+1}) \geq (1-2\gamma) H,
\end{equation*}
completing the proof of the gap estimate. To get the fidelity bound
(\ref{eqn:fidelity}), label the eigenvectors of $H$ by $\ket{E_0} =
\ket\Psi, \ket{E_1}, \ldots \ket{E_{d^n-1}}.$ We know that the
eigenvalue $E_0$ equals zero. Since invertibility of the $\Gamma_j$'s
guarantees uniqueness of the ground-state
\cite{Fannes1992,Perez-Garcia2007}, the bound on the gap ensures that
all other eigenvalues fulfill $E_i\geq (1-2\gamma)$. Thus
\begin{eqnarray*}
	&&
	\sum_j \left(\Tr[h_j\sigma_j] + \epsilon_j\right) 
	\geq
	\Tr[H\rho] \\
	&=& \sum_i E_i \bra{E_i}\rho\ket{E_i} \geq
	(1-2\gamma)(1-\bra{\Psi}\rho\ket{\Psi}).
\end{eqnarray*}
Finally, we remark that based on the ideas in \cite{Fannes1992}, it
can be shown that $\gamma\to0$ exponentially fast as $k$ is increased. 

% -----------------------------------------------------------------------------------------------------------%
\begin{figure}[t!]
\begin{center}
\includegraphics[scale=.57]{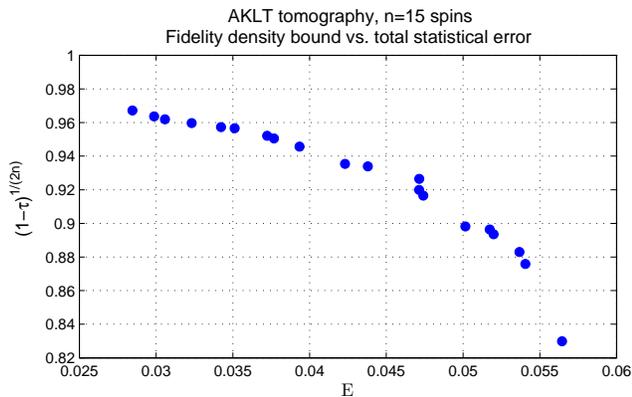}
\caption{Tomography of an AKLT state, a particular state of $n$ spin-1 particles with an efficient matrix product description.  The reconstruction method uses the DMRG method discussed in the main text for nearest-neighbor tomography only, and computes the error parameter $\tau$ by using the DMRG estimate of the gap.  The x-axis is the total error, $E = \sum_j \epsilon_j$, and the y-axis is the lower bound on the fidelity density $F^{1/n}$.  The true fidelity of the recovered matrix product description was near 1 in all cases, suggesting that the method for finding the state performs much better than what the bound allows us to certify.  The bound can be significantly improved if next-to-nearest-neighbor correlations are included.  \vspace{-20pt}}
\label{F:AKLT}
\end{center}
\end{figure}
% -----------------------------------------------------------------------------------------------------------%

\paragraph{Degeneracy and singular cases.} While the certification algorithm above works for ``most'' MPS states,
it may fail in certain singular cases. The simplest example is
given by the family of GHZ-type states $\frac1{\sqrt 2}(\ket{0,
\ldots, 0} + e^{i\phi} \ket{1,\ldots, 1})$. Since any local reduced
density matrix is independent of $\phi$, it is \emph{impossible} to
distinguish the members of that family based on local information
alone. (Indeed, for these states the $\Gamma_j$-operators defined
above will not be invertible -- so that an error would have been
declared). However, one may verify that the simple ``string operator''
$\sigma_x \otimes\ldots\otimes \sigma_x$ has expectation value
$\cos\phi$ for the states above. Thus, it is reasonable to expect that
H-MPS can be extended, enabling it to learn (and certify) any MPS with
bounded bond dimension.

It turns out that this possible, as we will briefly sketch now.
Indeed, if the algorithm described above reports an error (i.e.\ if
the $\Gamma_j$'s fail to be invertible), one would proceed as follows.
Employing the Quantum 2-SAT Method, compute an MPS representation of
the projector onto
the ground space of the parent Hamiltonian of the estimate. It may
happen that the dimension of this space grows exponentially as more
sites are taken into account. In this case, H-MPS will have definitely
failed (in the limit of vanishing noise, this only happens if no MPS
description with given bond dimension exists). However, in many
relevant situations, the dimension will saturate at a finite value.
Re-examining the gap estimate above shows that (unlike the original
proof in \cite{Fannes1992}), it remains valid for models with a
degenerate ground state space. Hence, at this point, we can certify
the overlap between the true state and a small subspace of the
exponentially large ambient space. 

It remains to be shown that even
this comparatively small ambiguity can be efficiently resolved.
Treating all special cases which may appear in the most general
situation is somewhat cumbersome and will be deferred to a future
publication. Here, we restrict attention to the most relevant case: a
two-fold degeneracy (this covers the GHZ and the W state). Formally,
we are facing the task of performing tomography in a copy of
$\mathbb{C}^2$, which has been embedded into $\mathbb{C}^{d^n}$. A
classic result states that any two orthogonal pure states of a
multi-partite system can be reliably distinguished using local
operations alone \cite{Walgate2000}. This holds true 
in particular for the
embedded versions of the eigenvectors of the three Pauli matrices
acting on $\mathbb{C}^2$. Therefore, the overlaps between a state in the
ground-space and this informationally complete set of vectors may be
found by local operations. We end the discussion by noting that the
procedure equally applies to the case where $\rho$ is far from a pure
state, as long as its support is contained in an MPS space.

\paragraph{Generalizations and Outlook.} The theory of MPS can be
generalized to higher-dimensional configurations, where the resulting
states are sometimes referred to as PEPS. We note that our gap
estimate does not crucially rely on the geometry of the neighborhood
relations. Therefore, conceptually, it should be relatively
straight-forward to generalize the basic techniques to these
situations. However, efficiently finding good PEPS approximations to
the data is likely to be extremely challenging.

While this Letter emphasized (approximately) pure states, tomography
of highly mixed states may actually be more pertinent in realistic
settings. If we drop the desire to certify the result, this task
proves relatively simple. Recall that a mixed state $\rho$ is called
\emph{finitely correlated} \cite{Fannes1992} if the linear space  of
operators
\begin{equation}\label{eqn:fcs}
	\left\{ \tr_{R} [\rho A_{R}] \,|\, A_{R} \in
	\mathcal{B}\left(R\right) \right\}
\end{equation}
has dimension smaller than some fixed constant $D^2$, where the
indices $L, R$ refer to a
bi-section of the chain into a left and a right part.
The symbol $\mathcal{B}(R)$ denotes the set of all operators acting on the
right hand side (r.h.s). Physically, the definition says that there
are only $D^2$ ways of modifying the state on the l.h.s.\ by
conditioning on an event taking place on the r.h.s.\ of the chain. It
has been shown in \cite{Fannes1992} that any state with that property
has an efficient representation, and once again it is natural to
ask whether this small set of parameters can be experimentally
obtained, and whether success can be certified without technical
assumptions. Under a mild invertibility assumption akin to
(\ref{eqn:gamma}), the answer to the first question turns out to be
affirmative. While details will be provided elsewhere, we briefly list
the main features of the algorithm: It is guaranteed to obtain a
faithful representation, without the need for any optimizations or
variational calculations; it yields a representation which may be used
to predict the expectation value of any (of the exponentially many)
factorizing observables, even though only (linearly many) local 
observables were measured to find the estimate. The MPS pure states
discussed above are included as a special case. On the downside, the
resulting ``matrix product operators'' are much harder to interpret;
there is no way of heralding errors if the assumptions are not met
and the resulting state will be non-negative only within the margins
of experimental accuracy.

We remark that our the results on mixed state tomography immediately
generalize to quantum process tomography. The analogue of a
non-factoring quantum state is here a channel with correlations over
many uses, while the analogue of the FCS representation are channels
with memory \cite{Kretschmann2005}. Obtaining the parameters of these
objects efficiently is, courtesy of the Choi-Jamiolkowski isomorphism,
just as straight-forward as learning mixed FCS.

\acknowledgments
% -----------------------------------------------------------------------------------------------------------%
\paragraph{Acknowledgments.} The authors thank 
R.~Blume-Kohout,
M.~Cramer, 
J.~Eisert,
O.~Landon-Cardinal,
Y.-K.~Liu, 
T.~Osborne,
M.~Plenio, 
D.~Poulin,
and
N.~Schuch
for discussions.  
SDB acknowledges the support of the Australian Research Council and the Perimeter Institute. 
STF and RS were supported by the Perimeter Institute for Theoretical Physics. Research at Perimeter is supported by the Government of Canada through Industry Canada and by the Province of Ontario through the Ministry of Research~\& Innovation. 
DG is glad to acknowledge support from the EU (CORNER).

% -----------------------------------------------------------------------------------------------------------%
% Bibliography
% -----------------------------------------------------------------------------------------------------------%

\bibliography{MPS}

\  \\

\paragraph{Note added.} After the research for this work had been completed \cite{qip2009}, the authors learned that two independent groups had worked along similar lines. 

Let us compare the present work with the recent work in Ref.~\cite{Cramer2010}.
Conceptually, the two contributions seem very close. Both make the
point that tomography of many-body quantum systems may not be as
impossible as it superficially seems. But the respective technical
focus is quite different.  While Ref.~\cite{Cramer2010} puts forward a new
algorithm for finding MPS approximations given measured data, we
emphasize methods for certifying that the estimate is actually
correct. Employing the MPS-SVT method of Ref.~\cite{Cramer2010} in step one of our H-MPS algorithm, it should be possible to seamlessly combine the two results.

Unrelatedly, the authors were recently made aware of the fact that a
further group was preparing results \cite{Poulin2010} on the problem
presented here.

\end{document}